%Paper: gr-qc/9306028
%From: tsg@iucaa.ernet.in (Ghosh)
%Date: Thu, 24 Jun 93 12:12:29 GMT

\magnification=\magstep1
\input paper
\voffset=-.5truein
\vsize=9truein
\baselineskip=14pt
\pageno=1
\pretolerance=10000
\def\n{\noindent}
\def\s{\smallskip}
\def\b{\bigskip}
\def\m{\medskip}
\def\c{\centerline}

\line{\hfil IUCAA-18/June'93}
\vskip 2 cm
\c{\mid GEOMETRIC PHASE IN VACUUM INSTABILITY:}
\c{\mid APPLICATIONS IN QUANTUM COSMOLOGY}
\vfill

\c{\it  DHURJATI    PRASAD    DATTA}
\b
\vfill

\c{Department of Mathematics, North Eastern Regional Institute of Science and
Technology,}
\c{Nirjuli  791 109, Itanagar, Arunachal Pradesh, INDIA$^{\dagger}$.}
\b

\c{and}
\b

\c{Inter-University Centre for Astronomy and Astrophysics,}
\c{Ganeshkhind, Post Bag 4, PUNE - 411 007, INDIA}
\vfill

\c{\bf ABSTRACT}

\vfill

\n Three different methods viz. i) a perturbative analysis of the
Schr\"odinger equation ii) abstract differential geometric method and
iii) a semiclassical reduction of the Wheeler-Dewitt equation, relating
Pancharatnam phase to vacuum instability are discussed. An improved
semiclassical
reduction is also shown to yield the correct zeroth order semicalssical
Einstein
equations with backreaction. This constitutes an extension of our earlier
discussions on the topic.

\b

\n PACS Nos:  03.65. - w  ; ~~04.60.  + n
\m

\n $\dagger$ Mailing address

\vfill\eject

\n The study of geometric phases $^{1-3}$ seems to offer important insights in
having a better understanding for a large class of physical problems. In
quantum field theory for instance the Berry phase appears to play a significant
role in elucidating
several conceptual issues relating to anomalies and associated problems. It is
shown$^{4, 5}$ that various gauge anomalies can be interpreted as due to a
non-trivial holonomy on the second quantized (chiral) fermion Hilbert
bundle over background static gauge fields. The non-trivial holonomy arises as
a
measure of topological obstructions in projecting the Fock vacuum in the
physical sector of the gauge manifold (static gauge fields mod local gauge
group). This in turn implies a loss of gauge invariance (global and non-abelian
anomalies) and/or an induced symmetry breaking (axial anomaly).

\s

\n Now the breakdown of the global $U(1)$ axial symmetry via an anomalous
divergence of the axial current induces axial baryon-lepton
non-conserving processes through the production of massless fermion
excitations$^6$.
Nelson and Alvarez - Gaume$^4$ have further shown that even the global and
non-abelian anomalies could be explained in terms of pair productions. Although
non-generic, the pair production occurs at the points of degeneracies of the
background field dependent Dirac hamiltonian, inducing a twist in the pertinent
Hilbert bundle.
\s
\n Recently some applications of Berry phase are also discussed$^{7,8}$ in the
semiclassical gravity in the framework of a minisuperspace cosmological
model. An improved Born-Oppenheimer analysis
in the Wheeler-Dewitt (WD) equation is shown to yield the correct zeroth order
semiclassical Einstein equations. The functional Schr\"odinger equation
describing
quantized matter fields in a background curved space is obtained at the next
order
of approximation. Further, the semiclassical backreaction of the matter fields
is shown to be determined by the U(1) Berry connection on the gravitational
sector of the
minisuperspace. An interesting consequence emerges in the Robertson-Walker
(RW) mini-superspace which is one-dimensional with trivial R-topology. The
relevant Hilbert bundle turns out also to be trivial, thereby reducing the
induced
Berry connection essentially to zero. As a consequence the WD equation
corresponding
to a gravitational action without a cosmological $\wedge$-term yields at the
semiclassical regime a matter Schr\"odinger equation essentially in the
Minkowsky space$^{8}$. However, for an action with a non-zero $\wedge$,
one gets zero matter equation in the DeSitter (DS) universe, although
zeroth order analysis does not yield a suitable backreaction. One however,
expects
a finite rate of particle production in the  DS background. It is, therefore,
of interest to see how the semiclassical Einstein equations with a reasonable
backreaction can be obtained through some modifications of the arguments in
Refs.[7,8].

\s

\n It is wellknown that the particle production in QFT in the presence of a
classical external field is associated with the vacuum decay which is
essentially
a non-perturbative effect. Under the influence of a time varying external
field the otherwise stable initial vacuum evolves into an admixture
of multiparticle states, thereby reducing the vacuum transition probability
amplitude to a value less than the initially normalized value one.
\s

\n Now, one naturally feels tempted to see if there is some intrinsic
relationship between the particle production through vacuum decay and the
particle production via
symmetry breaking due to anomaly. It is therefore of interest to look for
a description of particle creation through vacuum decay in the language of the
geometric phase. This will also enable one with important insights as to how
the modifications in the Born-Oppenheimer analysis is to be incorporated to get
a consistent set of  semiclassical Einstein equations.

\s

\n The motivation of the present paper is exactly this. We discuss some
wellknown examples of vacuum instability and show how a geometric phase can be
associated with the
decay width of the state. In sect. 2 we show in the context of a quantum
mechanical decay model that the decay width $\Gamma$ is related to the
Pancharatnam phase $^3$ between the initial and the final states. Pancharatnam
phase is a generalized geometric phase which may be obtained even for a
non-unitary, non-cycle evolution.
In particular, Pancharatnam phase may be non-zero even for a case where Berry
phase is zero or not sensible . Our mehtod uses a perturbative argument
although
the result is exact and non-perturbative. The result also agrees with more
abstract formulations$^5$ of the geometric phase (and anomaly). However, we
discuss this result
here as a prelude to our main result (Sect. 3). (The author is however unaware
of any prior explicit discussion of this example in literature.) In sect. 3, we
present
an extension of our earlier derivation$^{7,8}$ of the semiclassical Einstein
equations. This frees the earlier discussions from the necessity of a cyclic
evolution
. We show that for a non-cyclic evolution the backreaction can be related to
the
Pancharatnam phase. For topologically trivial minisuperspace where Berry phase
is
zero, this yields a set of semiclassical equations which describe gravity
induced instabilities in the matter Fock vacuum. The method also offers another
proof for the
formula  relating the Pancharatnam phase and the vacuum decay width.

\b

\n{\bf 2. VACUUM INSTABILITY IN QUANTUM MECHANICS}
\s

\n We consider the quantum mechanical decay of the ground state in the hump
potential

$$V = x^2 - \lambda x^4 ~~, \lambda > 0 \eqno(1) $$

\n This potential has a `bounce' solution with a single negative mode in the
Euclidean time $t_E = -it$. The standard instanton calculation$^9$ yields the
vacuum - vacuum amplitude.

$$ \langle f |i \rangle \equiv \int {\cal D}x \exp \left( -i \int^T_0 \left({1
\over 2} \dot x^2 - V \right)  dx \right) \simeq e^{-\Gamma T} \eqno(2) $$

\n where

$$\Gamma = |K| \sqrt S_0  ~~e^{ -S_0 }\eqno(3) $$

\n is the decay width of the state, $S_0$  the Euclidean bounce action and $K$
is a constant determinant factor. (By a suitable redefinition we absorb
the  harmonic oscillator ground state energy in the potential). The essential
feature of the expresion (2) is that the ground state energy of the
corresponding hamiltonian which is defined via a suitable analytic continuation
$-\lambda \rightarrow
\lambda \bar e^{i\pi} $ picks up a small imaginary part $\Gamma$ signalling the
instability. In the instaton calculation this is taken care of by the
negative mode in the bounce solution. Moreover, the basic object being the
transition probability amplitude, the inquiry into the existence of an
extra phase was not necessary in the standard discussion of the problem.
However, we are here primarily interested in calculating the non-trivial phase
of $\langle f|i \rangle$, if any.

\s

\n For this purpose we use an adiabatic perturbation method to analyse the
issue. Let us denote the relevant hamiltonian by $H(\lambda)$ and the
corresponding ground state $\psi
(\lambda)$. Introduce a Euclidean parameter $\tau$ periodic in $0 \leq  \tau
\leq 2 $ and denote by $\lambda_{\tau} = \lambda(\tau)$ a slowly varying
periodic function so that $\lambda (0)= 0$, $\lambda (1) =  \lambda$. By slowly
varying we mean
$\lambda (\tau) \simeq \lambda $ almost everywhere in [0, 1].

\n We now write

$$ \psi = e^{- \Gamma t} \phi  \eqno(4) $$

\n in the real time Schr\"odinger equation

$$i {\partial \over \partial t } \psi = H \psi  \eqno(5) $$

\n so that

$$H (\lambda) \phi (\lambda) = i \Gamma \phi (\lambda) \eqno(6) $$

\n As stated already the energy is pure imaginary due to the analytic
continuation $-\lambda \rightarrow \lambda \bar e^{ i\pi}.$

\n  The important fact to note is that equation(6) can as well be obtained by
treating
$\lambda_{\tau}$ perturbatively via the Euclidean equation

$$ H (\lambda_{\tau}) \chi_{\tau} = {\partial \over \partial \tau} \chi_{\tau}
\eqno(7) $$
\n The ansatz $$\phi_{\tau} = \bar e ^{ i \int^{\tau}_0 \Gamma_{\tau} d_{\tau}}
\chi_{\tau} \eqno(8) $$

\n then yields in the limit $\tau \rightarrow 1$, eqn.(6). The mechanism
however defines a parallel transport which generates a phase $\Gamma$
(for almost constant $\tau$-dependence in $\Gamma_{\tau}$ for the state
$\phi \left( \equiv \phi_1 \right)$. Eqn.(4) then gives the intended phase
relation$^{\dag}$

$$\psi = e^{- \Gamma t} e^{-i \Gamma} \chi \eqno(9) $$

\n Note that the states $\psi$ and $\chi$ belong to different rays. The
perturbatively generated phase $\Gamma$ between them is by definition the
Pancharatnam phase$^3$ which signals an induced twist in the line bundle of
states due to the perturbing potential $\lambda x^4$.

\s

\n Further insights into the phase relation (9) can be obtianed by letting
$\tau$ to make a complete circuit in $0 \leq \tau \leq 2 $.
After a complete cycle through the classically forbidden region the final
oscillator ground state $\psi (2)$ returns to the initial oscillator ground
state
$\psi(0)$ with however, an irreducible  phase $\left( - 2 \Gamma \right),$

$$ \psi (2) = e ^{-i2\Gamma} \psi (0) \eqno(10) $$

\n Although both the states are stable, the phase $2 \Gamma$ carries an imprint
of the twists in the perturbed line bundle . (The amplitude of $\psi$ in eqn.
(7) drops out since   $\Gamma_{\tau} \rightarrow 0 $ ~ as $ \tau \rightarrow 2
$). Stretching the analogy too far, in the
field theory language,  the vacuum decays with associated particle creation in
the first half of the circuit $0 \leq \tau \leq 1 $. However, in the other half
$1 \leq \tau \leq 2 $ the annihilation of particles occurs restoring the
initial vacuum. The whole process however leaves an imprint in the form of a
non-trivial phase shift indicating particle creation (annihilation) in the
intermediate stages. In the most of the physical situations though the two way
processes cannot be realized leading to genuine particle production. In the
case of gauge theories with chiral fermions the above cyclic process appears to
occur; the final irreducible phase indicates the absence of a global symmetry/
gauge invariance.
\s

\n We also note that the introduction of the Euclidean parameter $\tau$
alongwith the analytic continuation in the Schr\"odinger eqn.(5) provides a
complex structure
$^5$ in the quantum system. No such natural complex structure is available in
the case of tunnelling between degenerate vacua. So one does not expect a
geometric phase in this case.

\s

\n It is comforting to see that the non-trivial  phase in eqns. (9) and (10)
can also be obtained from a more abstract formalism$^5$. The present quantum
system actually corresponds to a hermitian holomorphic bundle over the
punctured complex plane $ C -
\{ 0 \} : z = \pi (t + i \tau)$.
\n The hermitian metric on this bundle is defined by the norm of the state
$\psi$

$$ \gamma \equiv \langle \psi | \psi \rangle = \exp \left( - {1 \over \pi}
\int \Gamma (dz + d \bar z) \right) \eqno(11) $$

\n The unique holomorphic connection corresponding to the metric (11) is given
by

$$ A = \gamma^{-1} \partial \gamma = -{1 \over \pi} \Gamma - {1 \over \pi}
\int { \partial \Gamma \over \partial z } d \bar z \eqno(12) $$

\n The corresponding curvature $ F = \bar \partial \gamma^{-1} \partial \gamma
$ vanishes identically because of the reality of $\Gamma$. However the
connection has a non-vanishing holonomy

$$ \oint A dz = - \oint \Gamma { dz \over \pi } - {1 \over \pi}\oint \int {d
\Gamma \over d z }
dz d \bar z \eqno(13 ) $$

\n The second  integral vanishes for a suitable choice of $\Gamma (\tau) $
$(\Rightarrow $
 vanishing
  residue of  $\partial \Gamma /\partial z~  at ~z = 0 $ ). Again for almost
constant $\Gamma$ along the cycle we have the desired phase $(-2 \Gamma )$.

\s

\n We close this section with the following remark. The general method of
holomorphic
line bundle can be applied to the QFT vacuum instabilities. As in the quantumm
mechanical model, the probability of pair creations in a finite volume is equal
to the Pancharatnam phase between the out and in vacuum states. An application
of the phase in gravity is discussed in the next section.

\b

\c{\bf 3. BACKREACTION AND PARTICLE PRODUCTION IN GRAVITY}

\s

\n We consider a minisuperspace gravity-matter system described by the
hamiltonian$^{7,8,10}$

$$\eqalign{ H &= H_g + H_m \cr
H_g &= - { 1 \over 2M} \bigtriangledown_g^2 + MV(g) \cr}\eqno(14) $$

\n where $H_g$ stands for the gravitational and $H_m$ for the matter
hamiltonian. We represent the matter fields by the symbol $\varphi$. The WD
equation assumes the form

$$\left( - { 1 \over 2M} \bigtriangledown_g ^2 + MV (g) + H_m \right) \Psi (g,
\varphi) = 0    \eqno(15) $$

\s

\n In Ref [7,8] , it is shown that an improved Born-Oppenheimer
approximation$^{11}$ with the inclusion of a non-trivial Berry phase yields the
effective gravitational equation

$$ \left[ - {1 \over 2M}Dg^2 + MV (g) + \langle \psi | H_m | \psi \rangle
\right]
\phi (g) = 0 \eqno(16)  $$

\n where

$$\Psi (g, \varphi) = \phi (g) \psi (g, \varphi ) \eqno(17) $$

\n $D_g = \bigtriangledown_g - iA$ denotes the covariant derivative due to
the induced $U(1)$  adiabatic connection

$$ A = i \langle \psi | \bigtriangledown_g \psi \rangle \eqno(18) $$

\n Further, using standard semiclassical analysis$^{10}$ around the expanding
WKB state $\phi(g) \sim \exp (-i S(g) )$ the curved space equation is obtained
in the Schr\"odinger picture at the order $0(M^0) :$

$$ i { d \over dt} \psi = -      H_m \psi \eqno(19) $$

\n where the `WKB time' $t$ is defined by

$$ { d \over dt} = \bigtriangledown_g S. \bigtriangledown_g \eqno(20) $$

\n The zeroth order Einstein equation is retrieved to order $0(M)$

$$ { 1 \over 2M} P^2    _{eff} + V(g)  + \langle \psi | H_m | \psi \rangle  =
0\eqno(21) $$

\n where $P_{eff}$ is the effective grvitational momentum.
We also note the relation

$$ \bigtriangledown_g ~S .  A = - \langle \psi | H_m | \psi \rangle \eqno(22)
$$

\n It thus follows that the backreaction in the form of an energy expectation
value  gets determined by the Berry connetion A. However for a simply
connected minisuperspace with flat geometry the connection A can be gauged away
$ A \equiv 0$, yielding instead of eqn.(21) the source-free equation$^8$

$$ { 1 \over 2M}    P_{free}^2 + V(g) = 0 .\eqno(23) $$

\n Thus the zeroth order backreaction cannot be obtained from the above
argument. For example, in a RW minisuperspace without a $\wedge$-term ,  $V(g)
= g^2 \left[ g =
{\rm scale factor} \right] $ and a self consistent solution of eqns. (15 )and
(23) is a
flat Minkowsky space obtained via a Euclidean-continuation $^8$. This means
that
the semiclassical reduction of WD equation for a Friedmann-like model yields
only a Minkowski space matter Schr\"odinger equation. However, for a non-zero
$\wedge$-term, the reduction yields a  DS space as a solution of eqn (23). The
matter eqn. (18)
is therefore a DS space Scr\"odinger equation. The exponential expansion of the
scale factor must therefore produce particles through DS vacuum instability
which, it is expected, should  backreact to gravity.

\s

\n One must therefore look for this backreaction  either at a higher order of
the
semiclassical approximation or by a modification of the Born-Oppenheimer scheme
applied to eqn.(15). We prefer the later to get a zeroth order backreaction
even for a
simply connected flat minisuperspace.

\s

\n We again start from eqn.(17)

$$ \Psi  ( g, \phi ) = \phi (g) \psi (g, \varphi) \eqno(24) $$

\n Choose two different values of $g~ : g_1  $ and $ g_f$
corresponding to two `instants' of quasiclassical evolution. Define, for
convenience,
$| \psi_i \rangle  = \psi (g_i  , \varphi ) $ the initial normalized Fock
vacuum :
$ \langle \psi_i | \psi_i \rangle = 1$.  We also assume that all excited states
in the initial Fock column are empty: $| \psi > < \psi_i = {\bf 1} $. Now
instead of projecting the whole WD eqn(15) on $| \psi_i \rangle$ itself we
choose to project on a final Fock state
$ | \psi_f \rangle = \psi ( g_f, \varphi )$:

$$ \left[ - {1 \over 2M} \left( \bigtriangledown_g - i A \right)^2 + MV +
{ \langle \psi_f| H_m | \psi_i \rangle \over \langle \psi_f | \psi_i \rangle}
\right]
 \phi  = 0 \eqno(25) $$

\n where A is given by

$$ A = i \langle \psi_i | \bigtriangledown_g \psi_i \rangle \eqno(26) $$

\n For an adiabatically evolving gravitational mode, A is the Berry
connection(18).
However, the derivation of eqn.(25) does not require the need of a cyclic
evolution and may be applied even to a non-cyclic case. Using the definition of
time eqn.(20) one gets

$$ \bigtriangledown_gS .   A = i \langle \psi_i | { d \over dt} \psi_i
\rangle\eqno(27)
$$

\n Further using the Schr\"odinger eqn(19) and the inverse of the time
definition $ { dg \over dt} = \bigtriangledown_g S$, one obtians the
Pancharatnam phase$^{11}$ for the transition $ | \psi_i \rangle \rightarrow |
\psi_f \rangle$ in the form

$$ \int_i^f A. dg = - \int^{t_f}_{t_i} { \langle f |H_m | i \rangle \over
\langle f | i \rangle }   dt \eqno(28) $$

\n A comment is in order here.

\item{i)} The unitarity condition $ | \psi > < \psi| =    {\bf 1}$ asserts that
the parallel transport of states is to be done along the horizontal
subspace$^{2,3}$ only. The condition for such a horizontal transport is $
\langle \psi | { d \over dt} \psi \rangle = 0 \Rightarrow A = 0$ in the
intermediate stage $ t_i \leq t < t_f $. However, the condition of horizontal
transport fails at $ t = t_f $. Consequently
a mixing of the initial Fock states is allowed in the final Fock vacuum
yielding a non-zero value for the phase integral (28).
\s
\n Continuing with the discussion of the phase integral (28) we note that the
instantaneous vacuum energy of the Hamiltonian $H_m$ in the presence of an
induced instability assumes a small imaginary part $ E_0 + ^i\Gamma$. Here
$E_0$ is a
possible non-zero energy due to vacuum polarization and $\Gamma$ is related to
the vacuum decay width. The integral in the r.h.s. of eqn.(28) has the formal
expression
$ \int^t_i (E_0 + i\Gamma )dt$. To get a real value one must evaluate the
integral for $\Gamma$ along the Euclidean time $\tau = it$. Further, in the
case
of a simply-connected flat minisuperspace the integral $\int E_0 dt$ can be
safely gauged away. Eqn.(28) together with the remark (i) therefore yield the
Pancharatnam phase associated with an instability :

$$ \int^f_i A. dg = - \Gamma \eqno(29) $$

\n which agrees with the phase obtained in sect. 2. Infact. eqn(29) is another
derivation of the fact that the Pancharatnam phase indicating an instability
has to be
pinned via a transport along a Euclidean time$^{13}$. The result is exact
(modulo
adiabatic condition) and appears to have a general validity.

\s

\n In the absence of an instability eqn(29) yields a vanishing Pancharatnam
phase. This agrees with the result of Ref[8] that for a Lorentzian evolution
emerging from a flat simply connected minisuperspacee the corresponding Berry
phase (infact, connection)
is zero. (For a curved minisuperspace the real time energy integral (28) might
yield a meaningful geometric phase. This issue will be taken up seperately).

\s

\n The semiclassical backreaction equation corresponding to eqn.(25) thus
assumes the
form

$$ { P_g^2 \over 2M} + MV(g) + { \langle f |H_m|i \rangle \over \langle f|i
\rangle}
= 0 \eqno(30) $$

\n where $P_g$ is the source-free gravitational momentum. Note, however, that
the gravitational component $ { P^2_g \over 2M} + MV(g)$ in eqn.(30) is of
order
$0(M)$ whereas that of the backreaction one order less $0(M^{\circ})$. Thus
a reasonable set of Einstein equations obtained via a semiclassical reduction
must assume the following iterative form :

$$ 0(M) : { P^2_{g_0} \over 2M} + M ~ V (g_{\circ}) = 0 \eqno(31a) $$

$$ 0(M^{\circ}) : i { d \over dt} \psi (g_0, \varphi) = H_m \psi
(g_0, \varphi) \eqno(31b) $$

\n and the backreaction eqn.(30) is obtained only as a 2nd order iterated
equation:

$$ { P_g^2 \over 2M} + MV (g) = R_e { \langle f|H_m|i \rangle_{g_0} \over
\langle f| i \rangle_{g_0} }, \eqno(31c) $$

\n the imaginary part in the r.h.s. being  exponentially small is neglected in
the adiabatic approximation.

\s

\n In the case of a closed RW minisuperspace with a pure Einstein-matter action
eqn.(31a) does not have a reasonable solution in the Loretzian sector (because
momentum $P_{g_0}$ becomes imaginary). However, the equation yields a flat
Euclidean
solution which via an analytic continuation implies in turn that the evolution
of matter is essentially desribed by a Minkowsky space Schr\"odinger equation.
In this case no gravitationally induced instability is possible and hence
eqn.(31c) reduces to eqn.(31a) [normal ordered vacuum energy expectation value
vanishes in Minkowsky
space ].
\s

\n However, for an action with a positive cosmological term, eqn. (31a) yields
a DS
 space as a solution. The matter eqn. (31b) is thus a DS  vacuum Schr\"odinger
equation which is supposed to produce particles (modulo technicalities
in defining appropriate Fock states) because of an instability$^{14}$ induced
by
the exponential expansion. Eqn.(31c) then describes a possible modification in
the  DS metric by an appropriate backreaction.

\s

\n Before closing the discussion we note that the relevant phase integral in
the presence of a Euclidean wormhole structure assumes the form

$$ \int^f_i A. dg = i \int^{\tau_f}_{ \tau_i} { \langle f | {\tilde H_m}
| i \rangle \over \langle f | i \rangle } d \tau \eqno(32 ) $$

\n  Here $\tau$ is a Euclidean time parametrizing the wormhole handle
and $\tilde{H_m}$ denotes an appropriate matter Hamiltonian. It is
wellknown$^{15}$
that the occurence of a wormhole needs a complex matter field.  Existence of a
non-trivial geometric phase  along a non-contractible wormhole handle  now
suggests that $\tilde{H_m}$ must
 be realizable as an anti-hermitian operator on a Euclidean Schr\"odinger
energy eigenstate. (This particular point has not been explicitly stated in
Ref. 7).

\s
\n{\bf 4. FINAL REMARKS}
\s
\n Three different methods are shown to yield the same phase relation between
the in and out vacua in the presence of an instability. It is clear that an
instability
occurs whenever the Hilbert bundle associated to a given quantum system  has a
natural complex structure. In the examples discussed here the complex structure
arises from the punctured complex plane of the analytically continued physical
time. Although
the moduli space of a punctured plane (which is topologically a torus) is
non-trivial
, the present discussion suggests that the Pancharatnam phase for an unstable
vacuum is insensitive to the class of inequivalent complex structures. In any
case, it is however of interest to study any possible relationship between  the
vacuum instability and the moduli space of a torus.
\s

\n The present discussion also suggests a general unambiguous method of
obtaining semiclassical Einstein equations from a fully quantized system: It is
however unclear the precise sense how the energy expectation values capture the
backreaction of the particles produced in the cosmological background. In any
case, it is of much interest to see how this argument applies to a more general
superspace. It is also of interest to
substantiate the general results discussed here by explicit calculations.

\b

\n{\bf ACKNOWLEDGEMENTS}
\s
\n It is a pleasure to thank Inter-University Centre for Astronomy and
Astrophysics, Pune for kind hospitality under its Associateship programme. I am
particularly thankful to Prof. J.V. Narlikar, Director, IUCAA, Prof. N.
Dadhich, Drs. V. Sahni and A. Kshirsagar for stimulating discussions. Thanks
are also due to Ms. Manjiri Mahabal for her help in  \TeX.

\vfill\eject

\c{\bf REFERENCES}
\s
\item{1.} M.V. Berry, Proc. R. Soc. Lond. A 392, 45 (1984);
B. Simon, Phys. Rev. Lett. 51 2167 (1983).
\s
\item{2.} Y.Aharonov and J.Anandan, Phys. Rev. Lett. 58, 1593 (1987)
\s
\item{3.} J. Samuel and R. Bhandari, Phys. Rev. Lett. 60, 2339 (1988);
N. Mukunda, Quantum Kinematic approach to the Geometric phase - I :
General Formalism, Preprint, Indian Institute of Science, Bangalore, 1992.
\s
\item{4.} A.J. Niemi and G.W. Semenoff, Phys. Rev. Lett. 55, 927 (1985);
H. Sonada, Nucl. Phys. B266, 410 (1986);
P. Nelson and L. Alvarez-Gaume, Commun. Math. Phys. 99, 103 (1985)
\s
\item{5.} M.J. Bowick and S.G. Rajeev, Nucl. Phys. B 296, 1007 (1988);
S.N. Solodukhin, Commutator gauge anomaly, bundle of vacua and particle
creation in two dimensions;
ICTP preprint IC/91/120
\s
\item{6.} C.G. Callan, R.F. Dashen and D.J. Gross, Phys. Lett. 63B, 334 (1976)
\s
\item{7.} D.P. Datta, Mod. Phys. Lett. A8, 191 (1993);
        ibid 8 , 601 (1993).
\s
\item{8.} D.P. Datta, Semiclassical backreaction from Wheeler-Dewitt equation,
             Submitted
\s
\item{9.}S. Coleman, `Uses of Instantons', in : `The whys of Subnuclear
 Physics ' Ed. A. Zichichi, Plenum Press (1979) New York
\s
\item{10.} T. Banks, Nucl. Phys. B 245, 332 (1985); J.B. Hartle, in Gravitation
in Astrophysics, ed. J.B. Hartle \& B. Carter (Plenum, 1986); T.P. Singh and T.
Padmanabhan, Ann. Phys. (N.Y.) 196,  296 (1989)
\s
\item{11.} F. Wilczek and A. Zee. Phys. Rev. Let. 52, 2111 (1984)
\s
\item{12.} For a beautiful exposition of the salient features of Pancharatnam
Phase, we refer to Samuel and Bhandari in Ref 3.
\s
\item{13.} A similar observation in a slightly different context is made
recently by T. Kashiwa, S. Nima and S. Sakoda, Ann. Phys. 220, 248 (1992)
\s
\item{14.} N. Myhrvold, Phys. Rev. D 28, 2439 (1983);
L.H. Ford, Phys. Rev. D 31, 710 (1985)
\s
\item{15.} R.C. Myers, Nucl. Phys. B 323, 225 (1989) and references therein.

\vskip 1 cm
\n $\dag$ The phase $-\Gamma$ is dimensionless. The period of the Euclidean
parameter
$\tau$ is determined by the intrinsic time scale of the problem fixed by the
harmonic oscillator groundstate energy. We also set $\hbar = 1 $.
\vfill

\end